\documentclass[reprint,amsmath,amssymb,superscriptaddress,nolongbibliography,aps,prl]{revtex4-2}

\usepackage{xr}
\usepackage[utf8]{inputenc}
\usepackage{amsmath}
\usepackage{amsmath,amssymb}
\usepackage{siunitx}
\usepackage[dvipsnames]{xcolor}
\usepackage{color,soul}
\setstcolor{red}
\sethlcolor{yellow}
\usepackage{notes2bib}
\usepackage[colorlinks=true]{hyperref}
\hypersetup{
    allcolors = {blue},
}

\newenvironment{eqs}%
{\begin{equation} \begin{aligned}}%
{\end{aligned} \end{equation} }
\newcommand{\beal}{\begin{eqs}}
\newcommand{\eal}{\end{eqs}}

\usepackage{graphicx}
\usepackage{dcolumn}
\usepackage{bm}
\usepackage{booktabs}
\usepackage{framed,enumitem}

\begin{document}
\preprint{APS/123-QED}


\title{Superlubric-Locked Transition of Twist Grain Boundaries in 3D Crystals}
\author{Jin Wang}
    \affiliation{International School for Advanced Studies (SISSA), Via Bonomea 265, 34136 Trieste, Italy}
    \affiliation{International Centre for Theoretical Physics (ICTP), Strada Costiera 11,34151 Trieste,Italy}

\author{Erio Tosatti}
  \email{tosatti@sissa.it}
  \affiliation{International School for Advanced Studies (SISSA), Via Bonomea 265, 34136 Trieste, Italy}
  \affiliation{International Centre for Theoretical Physics (ICTP), Strada Costiera 11,34151 Trieste,Italy}


\begin{abstract}

Properties of twist grain boundaries (TGB), long known structurally but  not tribologically,   are simulated  under sliding and load, with Au(111) our test case. The load-free TGB moir\'e is smooth and superlubric at incommensurate twists.
Strikingly, load provokes a first-order structural transformation,  where  the highest energy moir\'e nodes are removed -- an Aubry-type transition for which  we provide a Landau theory  and a twist-load phase diagram. 
Upon frictional sliding, the transformation causes a superlubric-locked transition, with a huge friction jump, and irreversible plastic flow.
The predicted phenomena are robust, also recovered in a Lennard-Jones lattice TGB, and not exclusive to gold or to metals.

\end{abstract}

\maketitle


The nanoscale behavior of crystal-crystal interfaces, ubiquitous in materials science and engineering \cite{Pashley.1965,Allain.natmater.2015,Yan.review.2024,Shi.chemrev.2020,Lu.science.2009}, is increasingly important in rapidly developing fields including nanoscience and twistronics \cite{Neto.science.2016,Sun.chemrev.2024}.
While most current studies focus on structural, energetic, and electronic properties in free or constrained mechanical equilibrium, less attention is devoted to non-equilibrium mechanical, dynamical and frictional behavior.
Specifically, structurally mismatched crystal interfaces raise the question, whether under shear stress their mutual sliding could be frictionless, also designated as superlubric \cite{Shinjo.surfsci.1993}, or alternatively pinned by a frictional barrier.
Understanding the conditions for superlubric sliding and its possible transition to a pinned state is of considerable significance, superlubricity being pursued as a holy grail by engineers who wish to reduce frictional losses, pinning conversely ensuring lateral mechanical stability.
A comprehensive understanding of the expected grip or lubricity of 3D materials with realistic populations of grain boundaries is also fundamental for advancing the design and performance of micro and nano devices, such as actuators, bearings, generators, and precision sensors \cite{Luo.nanoenergy.2021,Huang.nc.2021,Huang.nc.2025}.

Bilayers of two-dimensional (2D) materials \cite{Hod.nature.2018,Wang.rmp.2024}, 2D/3D interfaces \cite{Dietzel.prl.2013,Wu.nanolett.2024}, and colloidal monolayers in incommensurate optical lattice potentials \cite{Bohlein.prl.2012,Bohlein.natmater.2012}, whose total free energy is independent of $(x,y)$ position offer exquisite 2D examples of superlubricity.
However, it was discovered long ago that incommensurability does not guarantee free sliding. Under the action of some external parameters, superlubricity may be lost through a so-called Aubry transition where sliders becomes spontaneously pinned, despite genuine incommensurability \cite{Brazda.prx.2018}.
At that transition, a minimal fraction of local atom-atom configurations with highest energy (whose proportion was called disorder parameter \cite{Coppersmith.prb.1983}) are eliminated, destroying the adiabatic continuity necessary for free sliding \cite{Aubry.1983}.
The basic question which we address here is whether superlubricity in the first place, and then superlubric-pinned transitions could occur in ordinary 3D metal-metal interfaces. We argue in this Letter that the answer is positive on both accounts.\\

Twist (111) grain boundaries (TGB), very common for example in fcc metals \cite{DeHosson.philomag.1990,Majid.philomag.1992,Zhang.asc.2005,Dai.actamater.2013}, represent an ideal test case in point.
The energy cost of a TGB as a function of twist angle $\theta$ is well known in classic literature to exhibit a non-analytic set of commensurate dips, overarched by a smooth, analytic  upper envelope  energy curve $E(\theta)$ \cite{DeHosson.philomag.1990,Dai.actamater.2014}.
That envelope identifies the cost of generically incommensurate TGBs, to which we shall restrict our attention.
At large twist angle $\theta$, the weak ``moir\'e'' distortions reflecting the twisted lattice mismatch at the TGB interface, are energetically irrelevant, leaving $E(\theta)$ essentially flat.
At low twists instead, $E(\theta)$ drops fast with $\theta$ towards its global minimum at $\theta$=0. That drop, analytically described in classic literature \cite{DeHosson.philomag.1990,Dai.actamater.2014} and the accompanying phenomena without and with external load, requires additional attention.
At low twist angle, the TGB moir\'e structure should generally undergo, as for example in twisted 2D bilayers \cite{Wang.rmp.2024}, a massive relaxation from an ideal rigid pattern to a ``reconstructed'' sparse network of narrow dislocations lines, enclosing large commensurate domains inside which the interlayer epitaxy is essentially perfect \cite{DeHosson.philomag.1990,Dai.actamater.2013,Long.cms.2020}.

Under shear stress, the sliding habit of an ideal incommensurate TGB interface of arbitrary incommensurate twist angle could in principle be frictionally superlubric -- or alternatively it could be pinned. 
Even in complete absence of any extrinsic defects or edges, pinning could arise if the moir\'e misfit dislocation network became spontaneously ingrained with the crystal lattice, giving rise to a Peierls-Nabarro barrier thus stifling free sliding.
In analogy with 1D systems \cite{Mukamel.pra.1989}, spontaneous pinning should be facilitated by two factors.
The small twist angle moir\'e reconstruction is one -- the pattern consisting of distant, narrow dislocations that only interact at isolated nodes. 
Another is a stronger lattice interlayer potential -- such as load could provide -- across the TGB interface. 
Lacking either theory or experiment, simulation can explore if and when superlubricity or pinning will prevail, with results that will call for fresh theory.\\

\begin{figure}[ht!]
\centering
\includegraphics[width=1.0\linewidth]{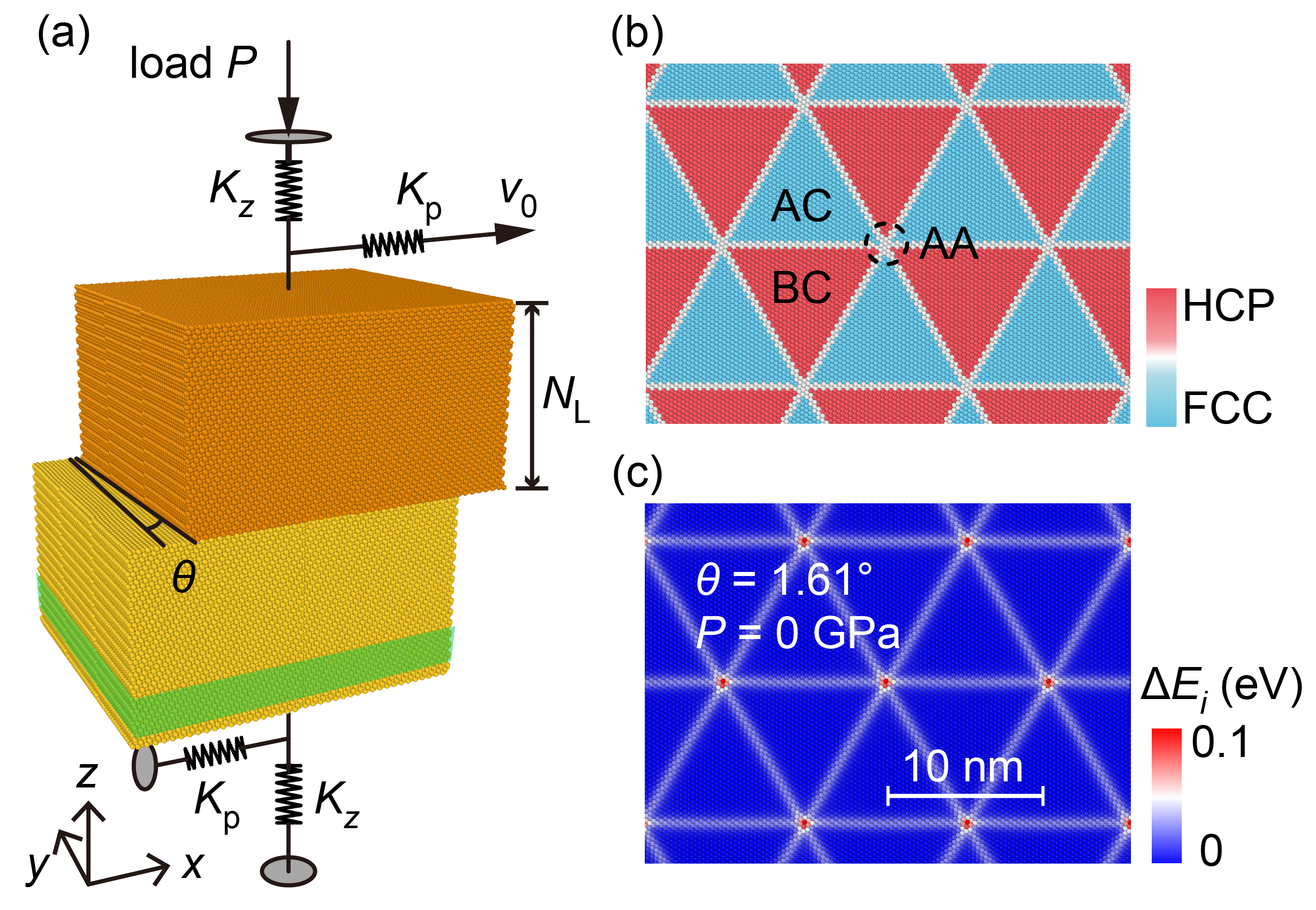}
\caption{Au(111) twist grain boundary (TGB).
(a) Simulation geometry, with adjustable twist $\theta$ and number of layers $N_L$. Periodic boundary conditions are used along  $(x, y)$.
A $z$-directed load and $x$-directed sliding force will be applied to the center-of-mass of the upper sliding block's top layer through separate springs.
The Langevin thermostated region in the lower block's bottom region is colored in green.
(b) Zero-load initial static structure showing the moir\'e network for a generic twist $\theta=1.61^{\circ}$, $N_L=48$. 
c. Per-atom potential energy difference $\Delta E_i$ above the $\theta$ =0 value, emphasizing the energy cost of the AA nodes, and to minor extent of the dislocations that join them.}
\label{fig:model} 
\end{figure}

Choosing gold as our test crystal, Au(111) TGBs were simulated with periodic boundary conditions along $x$ and $y$ directions for a set of small twist angles $\theta$. We used the LAMMPS code \cite{Plimpton.jcp.1995,Thompson.cpc.2022} with interatomic interactions described by an embedded-atom potential \cite{Foiles.prb.1986}.
The simulated TGBs lie between two 3D crystalline blocks, each a large number $N_L$ of (111) layers thick, as in Fig.~\ref{fig:model}(a).
The resulting interfacial stacking and per-atom energy of the optimized small-twist TGB ($T=0$) are pictured in Fig.~\ref{fig:model}(b-c). 
At small twist, the moir\'e network is indeed ``reconstructed", with narrow dislocation lines crossing at the AA nodes -- the highest interfacial energy spots \cite{Yoo.NatMater.2019,Kazmierczak.NatMater.2021}.
Relative to the unrelaxed moir\'e, the commensurate AC and BC regions expanded, whereas the AA and dislocation regions, where the overall twist of the interface is concentrated, contracted.
We studied the TGB structure both quasi-statically at $T=0$  and at $T=300$~K by Langevin thermostating 6 layers (Fig.~\ref{fig:model}a) in the lower block's bottom region \cite{Benassi.prb.2010}.
Normal load $P$ was applied through a rigid plate, connected to the center-of-mass of the top layer of the upper (slider) block via a $z$-directed spring of stiffness $K_z=N k_i$ ($N$ being  the atom number per layer, and $k_i=0.1$~N/m).
\footnote{This indirect loading mimics the elastic effects from the real semi-infinite body. For slabs of  significant thickness ($N_L=48$ was used for best results), the difference between indirect and direct loading (i.e., adding forces directly to top layer atoms) is minimal.}
Proceeding by steps, we explored first the  moir\'e structure and its changes with load at $T$ =0, then at $T=300$ K, where finally we carried out sliding simulations by dragging the top layer.

\begin{figure}[hb!]
\centering
\includegraphics[width=1.0\linewidth]{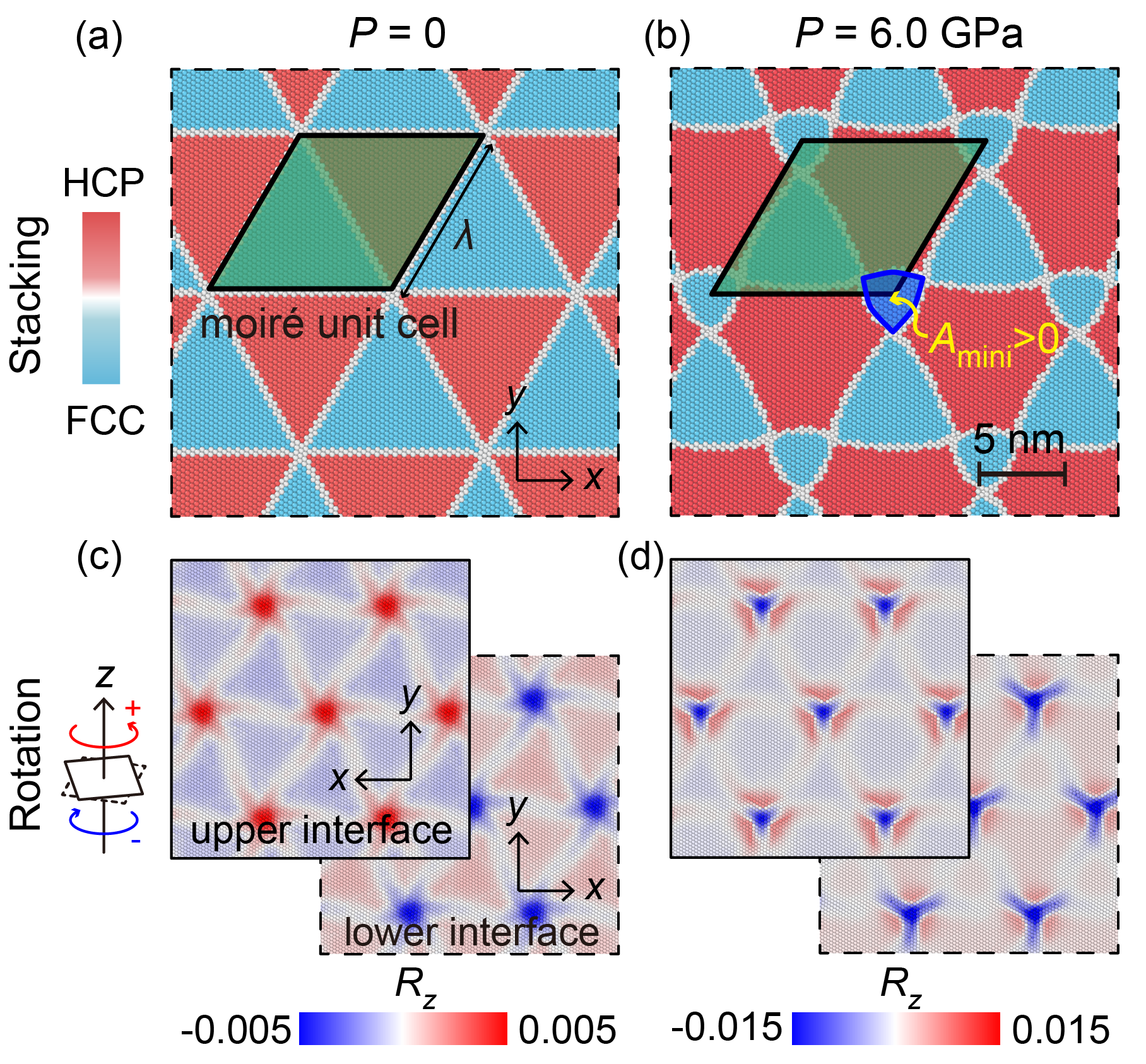}
\caption{Stacking (a,b) and local rotation (c,d) of a quasi-statically relaxed Au(111) TGB ($\theta=1.61^{\circ}$) under increasing load, $P=0$ (a,c) and 6.0 GPa (b,d). 
The blue and red colored regions represent fcc and hcp stacking respectively.
One of the mini domains (MDs) that replace the AA nodes is highlighted in deep blue in (b).
Their presence in the high load state makes the two facing layers totally different.
The local rotation tensor $R_z$ at the GB interface is measured with respect to the unrelaxed twist structure. The two facing interfaces are marked with solid and dashed borders respectively.}
\label{fig:structure}
\end{figure}

Quasi-static loading simulations were carried out with normal load raised from 0 to $10$~GPa in 0.1 GPa steps, allowing the $(x,y)$ slab size to expand so as to preserve zero in-plane overall stress.
The TGB reconstructed moir\'e network stood unscathed up to a twist-dependent instability pressure (e.g., $P\approx 6$ GPa for $\theta$ = 1.61$^\circ$), where an unexpected spontaneous structural transition occurred.
In the new structure the AA nodes (Fig.~\ref{fig:structure}a) were suddenly replaced by small but finite size commensurate mini-domains (MDs) of area $A$ (Fig.~\ref{fig:structure}b), inside which the fcc interlayer matching is perfect.
The dislocation network, previously confined at the 2D twisted interface, now reached out of plane -- the dislocation lines around the MDs invading the layer below the interface, as shown in Supplemental Material (SM-I) Fig. S1.
The cancellation of the highest local energy nodes is suggestive of a 3D-mutated version of the classic Aubry pinning transition of the 1D incommensurate Frenkel-Kontorova (FK) model \cite{Aubry.1983}, and of 2D twisted layers \cite{Mandelli.prb.2017, Brazda.prx.2018}.
A distinctive character of that transition is the ``disorder parameter" \cite{Coppersmith.prb.1983}, in our case defined by the MD linear size $Q \sim  A^{1/2}$, measuring the extent to which the costly AA nodes are spontaneously eliminated.
Unlike the 1D FK case, but similar to 2D colloidal monolayers \cite{Mandelli.prl.2015, Mandelli.prb.2017, Brazda.prx.2018}, the transition is  heavily structural and first-order, with a jump of all TGB properties and a strongly hysteretic behavior as a function of loading/unloading. 
An unrealistically high load is needed for the distortion to appear in quasi-static loading ($T=0$), whereas quasi-static unloading never reverts the structure back to undistorted, down to $P=0$ (Fig.~S2 in SM-II).
At the structural transition, the TGB symmetry also changed. 
The two facing layers initially equivalent by symmetry turned inequivalent after the distortion.
Specifically, the zero-load antisymmetry of the local rotation around the AA nodes in the two facing layers (represented by the $xy$-component of the rotation tensor, $R_z$) is lost once the commensurate MDs form (Fig.~\ref{fig:structure}c-d).
At the lowest twists studied, another distortion appears, only slightly higher in energy, where the MDs alternatively form in upper and lower layers. Properties of this ``up-down" distorted state, described in SM-III are not very different and are not pursued further here.\\

\begin{figure}[ht!]
\centering
\includegraphics[width=1.0\linewidth]{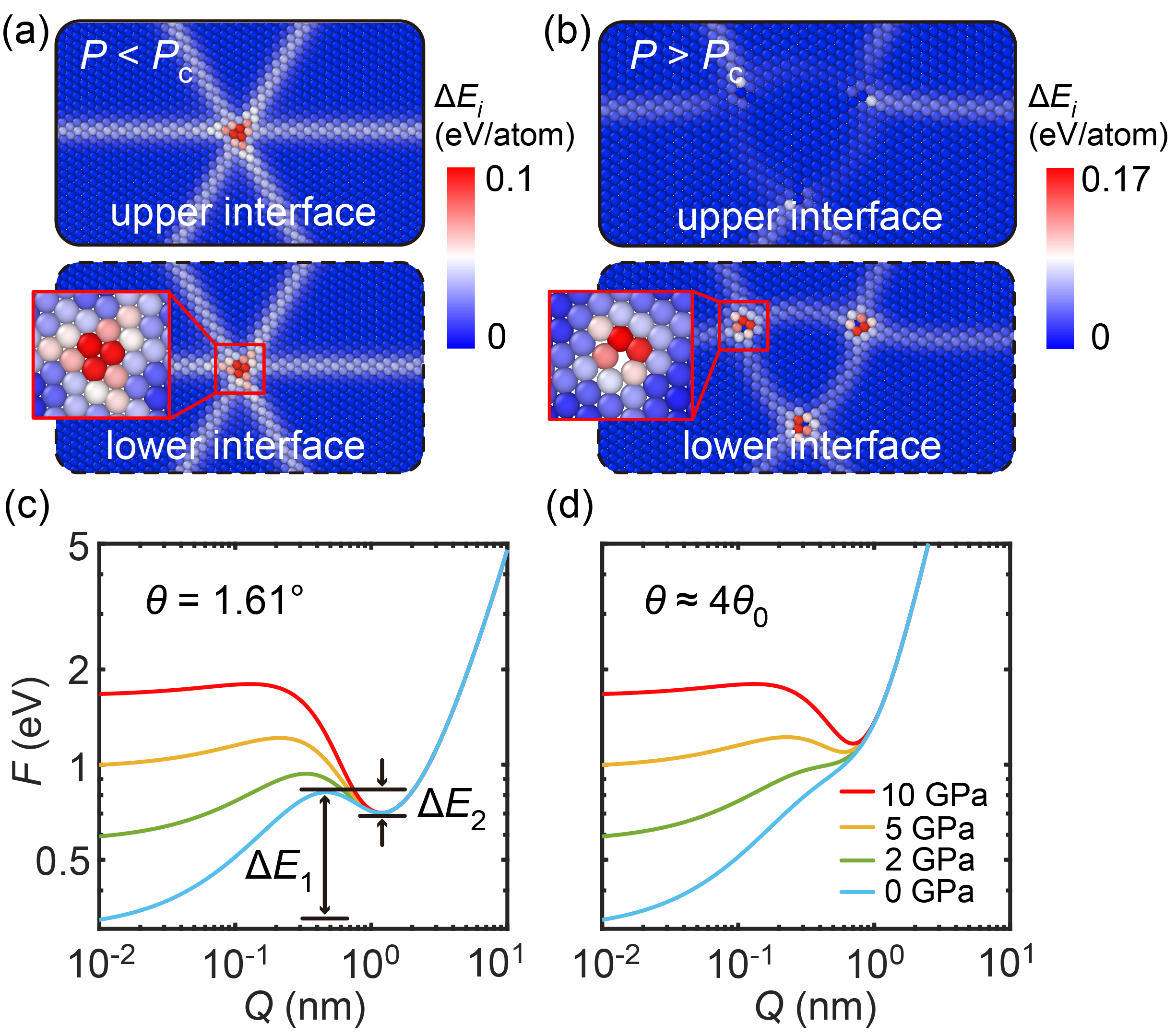}
\caption{Mechanism of the load-induced transition.
(a-b). Energy map near the AA nodes for $T=0$, $\theta=1.61^{\circ}$ below and above the transition.
Note the asymmetry between the two layers interfaces and the ``vacancies" at the MD corner for $P>P_c$.
(c-d). Free energy $F$ per moir\'e cell, Eq.~(1), versus disorder parameter $Q$ and load, at a small twist $\theta=1.61^{\circ}$ (c), and at the critical twist $\theta \approx 4\theta_0$ (d).}
\label{fig:mechanism} 
\end{figure}

The transition  mechanism seems transparent: the high AA node energy (Fig.~\ref{fig:mechanism}a) increases further under load $P$, eventually surpassing the limiting cost (both in-plane and out-of-plane) required to form the mini-domain in its place.
The mutual repulsion among each of the three MD corners (Fig.~\ref{fig:mechanism}b), the MD boundary formation, and the stress connected with the $\theta$ rotation in a 3D volume above the MD in turn conspire to determine the MD size. 
Together, these terms yield a mean-field theory, adequate in a first order transition where fluctuations are not crucial.
The extended Landau-type free energy per moir\'e cell can be written as a function of $Q$ as
\begin{equation}
\begin{aligned}  
 F(Q) &= F_1~f(Q/a_0) + F_2~[1-f(Q/a_0)] \\
 F_1 &=3 E_\mathrm{AA}\\
 F_2 &=3 E_\mathrm{vac} a Q^{-1} + \frac{\tau}{2} a^2 Q + \frac{G\theta^2}{2} a Q^{2}
\end{aligned}
\end{equation}
where $F_1$ and $F_2$ describe the two competing states, undistorted and distorted.
Here $E_\mathrm{AA}$ is the excess energy per atom of the three central AA atoms,
$E_\mathrm{vac}$ is a corner vacancy formation energy, $\tau$ is the shear strength, $G$ is the shear modulus, and $f$ is a cutoff function that can be taken as $f(x) = e^{-x^2}$.
The undistorted $Q$-independent AA node energy is load sensitive approximately as
\begin{equation}
    E_\mathrm{AA} (P) \approx E_\mathrm{0} + \alpha P a^3
\end{equation}
where $\alpha$ is a dimensionless parameter. Conversely, the distorted state energy is load independent, while its terms constitute a higher order polynomial that has a $Q > 0$ minimum.

With $E_\mathrm{0}=0.1$ eV, $\alpha = 0.3$, $a_0=3a/2$, $E_\mathrm{vac}= 0.5$ eV, $\tau = 1$ GPa and $G=30$ GPa, reasonable values for gold, the resulting relationship between free energy $F$, MD size $Q$, and load is qualitatively illustrated in Fig.~\ref{fig:mechanism}(c-d).
At zero load, the TGB  remains in the lowest energy $Q=0$ state.
As load increases, that energy rises, and the load independent minimum at finite $Q$ eventually prevails at a nominal transition load $P_t$, across a barrier of about 0.5 eV, reaching
\begin{equation}
   Q_0 \sim (\frac{6E_\mathrm{vac}}{\tau a})^{1/2}
\end{equation}
whence $P_t \sim  (\sqrt{2E_\mathrm{vac} \tau a^3/3} -E_0 )/(\alpha a^3)$. These approximations hold at small twist and $T=0$.
At larger $\theta>\theta_0$, where $\theta_0 = \frac{1}{3}[\tau^3 a^3/(2 E_\mathrm{vac} G^2)]^{1/4} \approx 2^{\circ}$ the $\theta^2$ term is important, leading to an increase of $P_t$ and a drop of the barrier height (Fig.~\ref{fig:mechanism}d).
Eventually, the barrier collapses and the first order distortion disappears altogether at a critical point of coordinates $(\theta_c, P_{tc})$, approximately at $\theta_c \approx 4 \theta_0$, 
and $P_{tc} \approx (3\sqrt{E_\mathrm{vac}\tau a^3/5}-E_0)/(\alpha a^3)$ (details in SM-IV).
In all small twist distorted states the MD size is less than half the moir\'e size, $Q/\lambda <1/2$.
As a consequence, the total TGB energy $E(\theta)$ well known in literature \cite{DeHosson.philomag.1990, Dai.actamater.2014} is modified below a critical twist angle $\theta_c$ and above the transition load $P_t$ by an additional term
\begin{equation}
    \Delta E(P, \theta) = F_2(Q_0, \theta) - F_1(P) <0
\end{equation}
For a low twist and high load such as  for example $\theta= 1.61^{\circ}$,  $P=5$ GPa,  $\Delta E = \sqrt{6E_\mathrm{vac} \tau a^3}-3E_\mathrm{AA}(P) \approx$ -0.3 eV/moir\'e cell (Fig.~\ref{fig:mechanism}c).\\

\begin{figure}[ht!]
\centering
\includegraphics[width=1.0\linewidth]{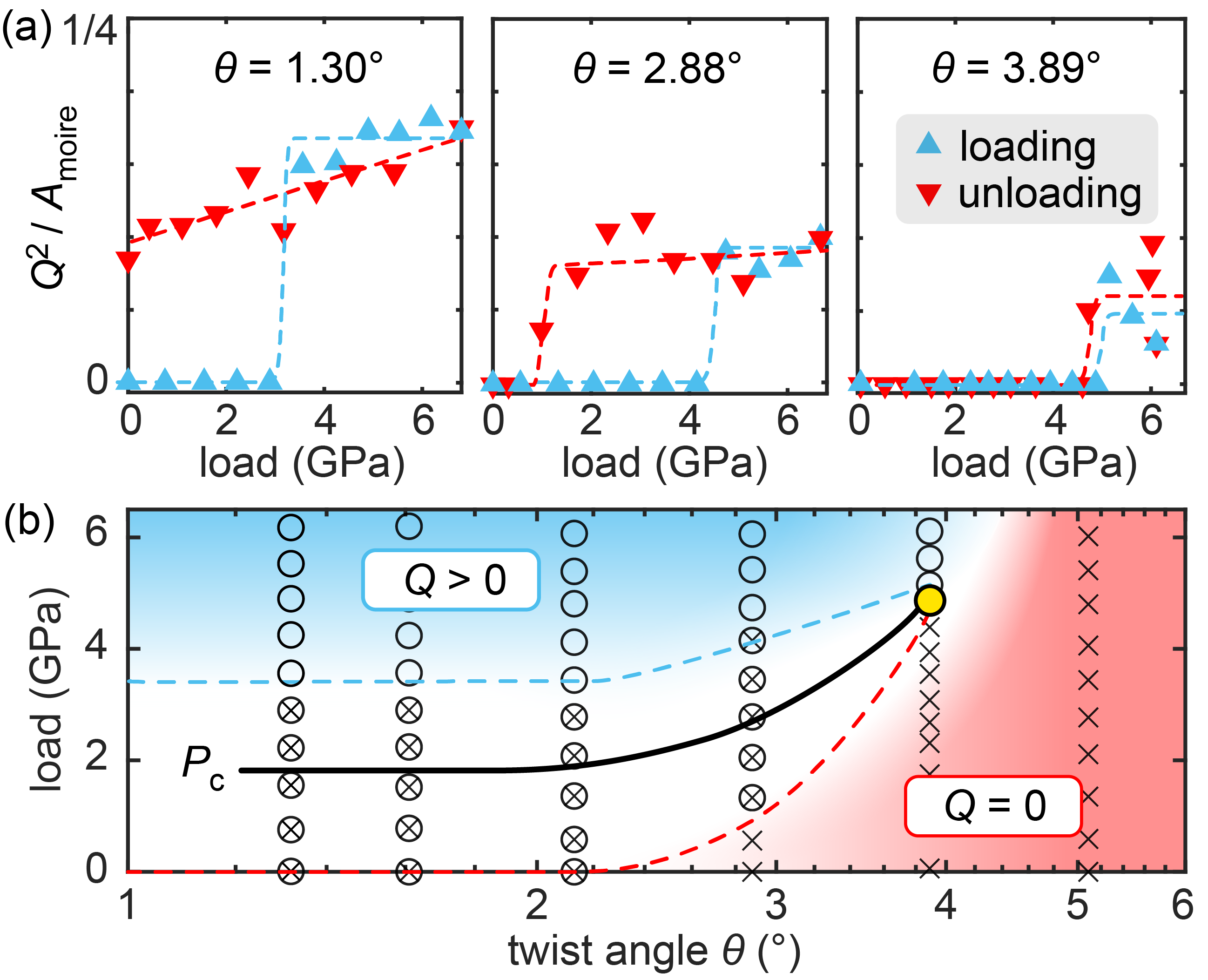}
\caption{(a) Load dependence of the disorder parameter $Q$ of the room temperature loading-unloading simulations of Au(111) TGBs at three representative twists.
(b) Phase diagram of $Q$ as a function of twist angle $\theta$ and load.
Circles and crosses represent cases where
$Q>0$ and $Q=0$ respectively during loading/unloading;
the crossed circles indicate cases where $Q=0$ during loading but $Q>0$ during unloading. Three phases are colored with blue, red, and white backgrounds.
Their boundaries are marked by blue and red dashed lines, with the median black line indicating the estimated first order transition load $P_c$ and the yellow dot marking the critical point.}
\label{fig:finiteT}
\end{figure}

\begin{figure}[ht!]
\centering
\includegraphics[width=0.96\linewidth]{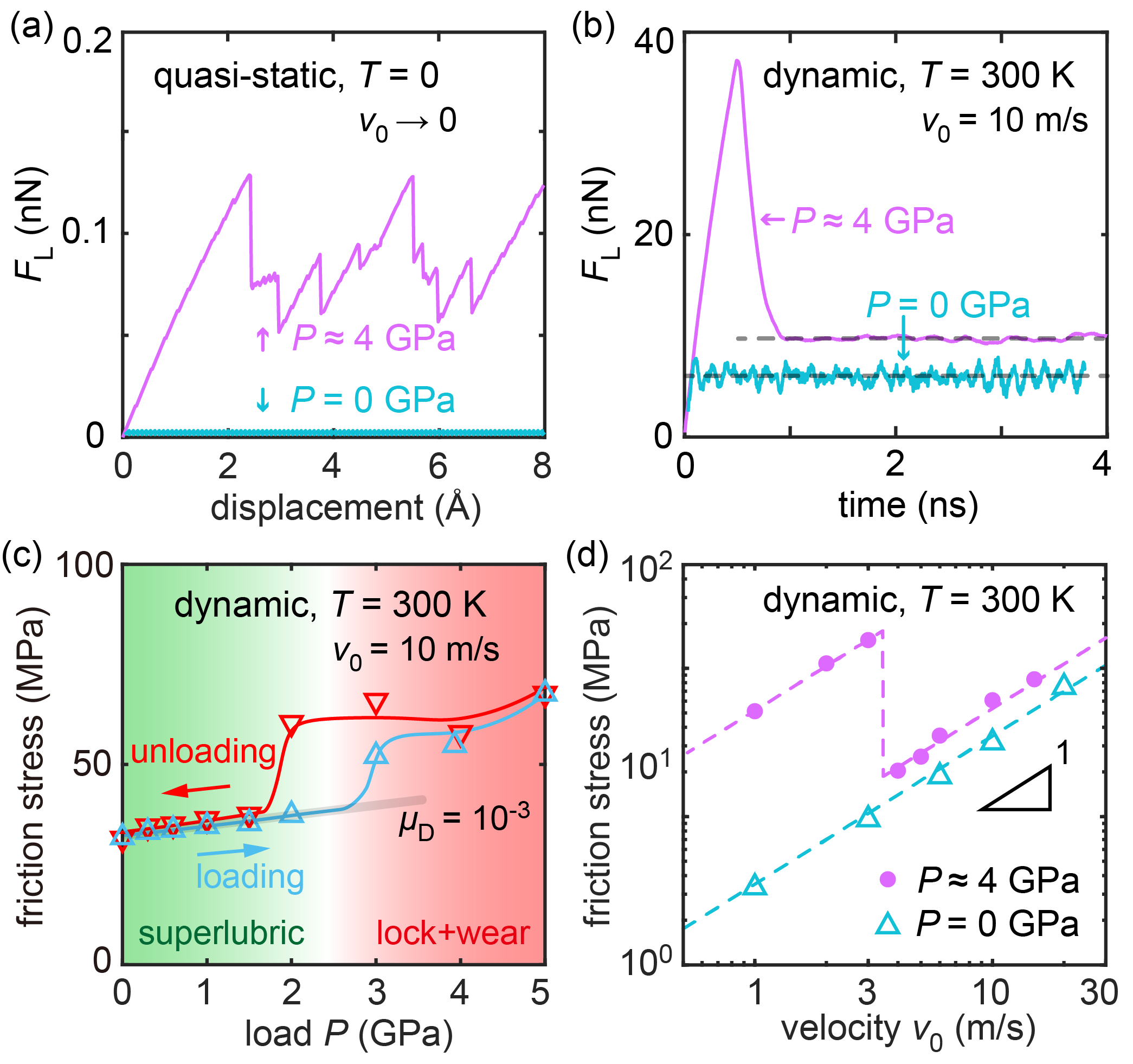}
\caption{Shear sliding of a small angle Au(111) TGB ($\theta=1.61^{\circ}$).
(a-b). Lateral (friction) force $F_L$ experienced by the dragging spring during sliding, both at zero load $P=0$ (blue) and at large load $P=4$~GPa (purple). 
(a) quasi-static friction; (b) kinetic friction at large velocity.
(c) Load dependence of friction for a loading/unloading cycle. The solid lines are guides to the eye.
(d) Velocity dependence of friction at zero load (undistorted TGB, blue) and at large load $P=4$~GPa (distorted TGB, purple) showing two plastic sliding regimes.
At high velocity the sliding occurs at the TGB interface; at low velocity sliding takes place elsewhere in the pristine lattice while the interface has become ``moir\'e cold-welded".}
\label{fig:friction} 
\end{figure}

Repeating simulations at $T_R$ =300 K instead of $T=0$, the distortive transition was found to persist.
The disorder parameter $Q$ as a function of load  and twist angle is shown in Fig.~\ref{fig:finiteT}(a).
For $\theta\leq 2^{\circ}$, the distortion on loading now occurs around 3 GPa instead of 6. Upon unloading, $Q$ diminishes while once again failing to  drop to 0 at $P=0$.
This hysteresis suggests that the barrier between distorted and undistorted states in small twist gold TGBs is still much larger than $k_\mathrm{B} T_R$  (Fig.~\ref{fig:mechanism}c).
For larger twists ($\theta$ from $3^{\circ}$ to $ 4^{\circ}$), $P_{t}(T_R)$ rises, with a narrower hysteresis cycle.
At $\theta \approx 4^{\circ}$ a critical point is finally reached, beyond which the distortion transition disappears altogether (Fig.~\ref{fig:finiteT}b).
The theoretical formulation given earlier at $T$ =0 remains applicable,  with normalized parameters, at room temperature. The phase diagram at $T_R$ of Fig.~\ref{fig:finiteT}(b) summarizes our results, where the black line in the middle of the hysteretic interval represents our prediction for $P_c$ and its dependence on twist angle. 
The width of that interval, outside which the TGB spontaneously recovers the equilibrium state -- undistorted for $P<P_t$,  distorted for $P>P_t$ -- should as usual be controllable, and reduced by manoeuvers such as thermal annealing or sliding.\\

As our final and crucial step, we studied the sliding behavior of Au(111) TGBs by pulling along $x$ the topmost layer through a spring of stiffness $K_\mathrm{p}=10$~N/m, with velocity $v_0$, and extracting the frictional force $F_L$.
An initial quasi-static pulling ($v_0\to0, T=0$) was applied to a virgin undistorted TGB  with $\theta =1.61^{\circ}$ to extract the static friction, subsequently followed by full room temperature dynamics simulations at increasing loads.
In the zero and low load undistorted TGB the static friction  was zero (Fig.~\ref{fig:friction}a). The kinetic sliding was also smooth (Fig.~\ref{fig:friction}b), with a low coefficient of friction, both ordinary and differential, on the order of $10^{-3}$  (Fig.~\ref{fig:friction}c) and with linear velocity scaling (Fig.~\ref{fig:friction}d).
Thus the sliding of the undistorted TGB is genuinely superlubric \cite{Wang.rmp.2024}.
Under high load the same TGB, now distorted above $P_t=2$ GPa, showed instead a large static friction, with a stick-slip-like quasi-static force pattern of approximate periodicity $\approx 2.9~\mathrm{\AA}$, the in-plane lattice constant (Fig.~\ref{fig:friction}a).
Dynamic sliding with $v_0=10$ m/s, $T=300$ K however produced an immediate severe TGB damage with plastic flow (Fig.~\ref{fig:friction}b) extending in various measures beyond the interface (details in SM-V).
Quantitatively, the steady-state kinetic friction magnitude also jumps first order-like, with a loading/unloading hysteresis cycle (Fig.~\ref{fig:friction}c) whose width also diminishes with decreasing velocity.
The load-induced structural transition thus led not just to wearless pinning, as could be expected in a standard 1D Aubry transition \cite{Vanossi.rmp.2013}, but to the outright TGB locking, with strong frictional plastic flow upon sliding.
At intermediate twist angles between zero and $\theta_c$ the locking even becomes, as exemplified in Fig.~\ref{fig:friction}d,  so strong to even surpass the regular (111) interplanar grip. 
In that case at low velocities, the sliding site switches remarkably from  the TGB plane, now effectively ``cold welded", out to crystalline 3D regions, with a strong friction jump.\\

Summing up, a first order structural transition is predicted in small-angle Au(111) TBGs under load, involving a symmetry change, where high-energy AA moir\'e nodes are replaced by commensurate  mini domains, representing an unprecedented form of 3D  Aubry transition. The transition persisted unchanged e.g., using quite different interatomic interactions \cite{Gupta.prb.1981,Rosato.phimag.1989}. 
It was in fact found in identical form in crystals very different from metals, such as the Lennard-Jones fcc lattice (details in SM-VI).
Despite a current lack of experimental data, this {\it de facto} ubiquity  suggests the Aubry distortion as a rather universal feature of loaded TGBs.
Tribologically, the load-free undistorted TGB is found to be structurally superlubric. Thus superlubricity is not confined to 2D materials interfaces, where its features have been intensely pursued so far \cite{Liu.prl.2012,Wu.prl.2024}.
It should instead be rife in generic 3D-3D twisted or otherwise incommensurate contacts, despite the presence similar intra- and inter-layer interactions -- of course as long as the stability and quality of these contacts will be properly controlled.
On the other hand, the load-distorted TGB is, despite overall incommensurability, frictionally locked  by a strong static friction. In turn, that arises by Aubry-type removal of the costly AA nodes, by what can be depicted as a cold welding at these localized spots in the moir\'e pattern.
That is in contrast with the 2D materials interfaces, where Aubry pinning apparently cannot be reached owing to excessively weak interlayer interaction \cite{Wang.aubry.tobepub}.
The frictional character of the distorted TGB  is  described as locked because, unlike the weakly pinned states described in literature \cite{Vanossi.rmp.2013,Buzio.acsami.2023}, here kinetic friction cannot proceed without damage and plastic flow. Interestingly, the frictional strength of the moir\'e cold-welded Au(111) TGB may even surpass that of the pristine lattice.
Our study in conclusion underscores several unique but hitherto ignored characteristics of real 3D material interfaces, systems which hold promise for broader applications of superlubricity and of strong load-induced locking.\\

\begin{acknowledgments}
Work carried out under ERC ULTRADISS Contract No. 834402. We are grateful for discussions with A. Silva and A. Vanossi.
\end{acknowledgments}

\bibliography{ref}

\end{document}



\title{Supplemental Material \\
Superlubric-Locked Transition in Twist Grain Boundaries}

\author{Jin Wang}
\affiliation{International School for Advanced Studies (SISSA), I-34136 Trieste, Italy}
\affiliation{International Centre for Theoretical Physics, I-34151 Trieste, Italy}

\author{Erio Tosatti}
\email{tosatti@sissa.it}
\affiliation{International School for Advanced Studies (SISSA), I-34136 Trieste, Italy}
\affiliation{International Centre for Theoretical Physics, I-34151 Trieste, Italy}

\maketitle

\tableofcontents
\newpage

\section{Dislocation networks}

\begin{figure*}[ht!]
\centering
\includegraphics[width=0.9\linewidth]{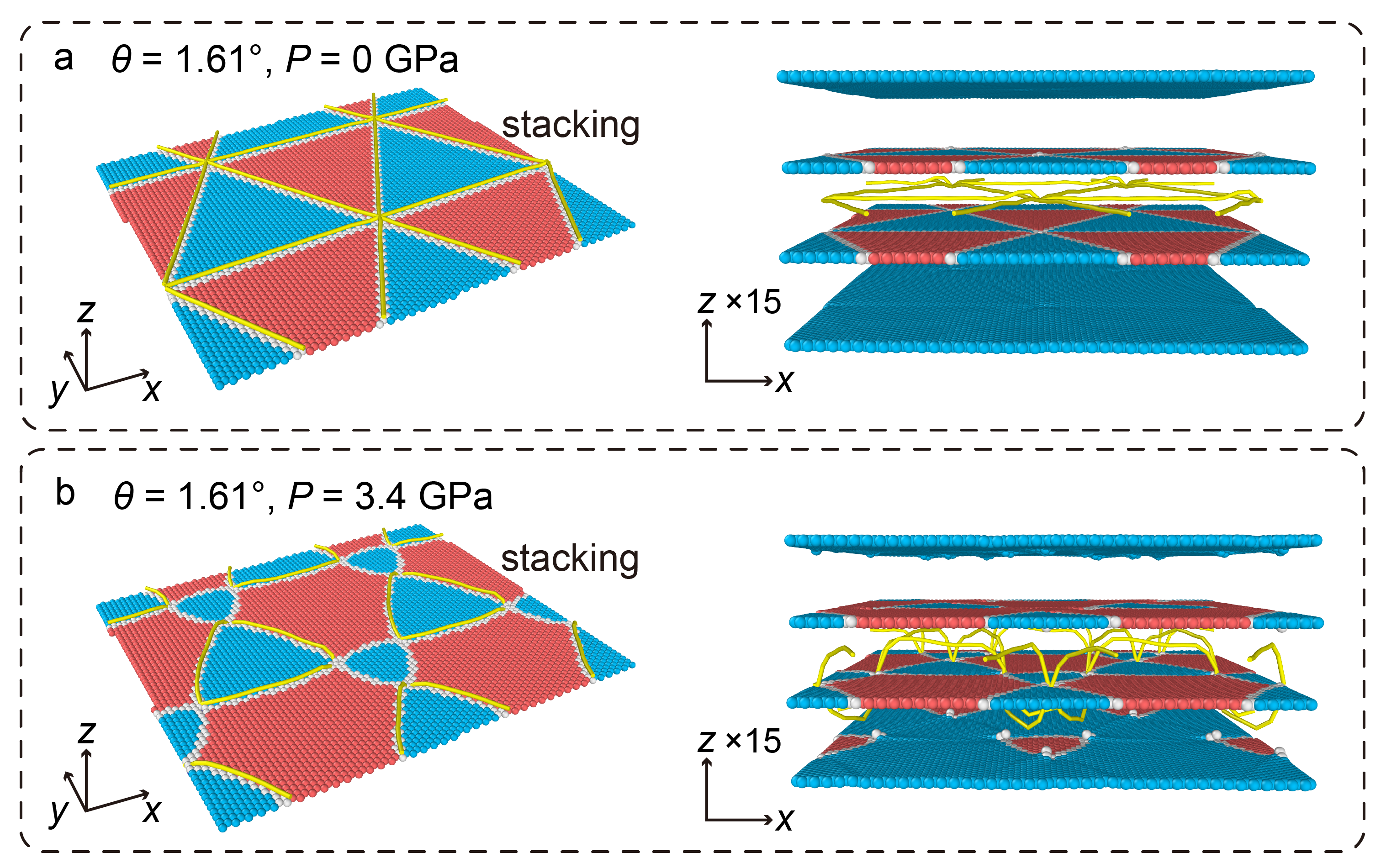}
\caption{Interfacial stacking and dislocation networks for a $\theta=1.61^{\circ}$ TGB under (a) zero normal load and (b) $P=3.4$ GPa.}
\label{figS:dislocation}
\end{figure*}

The interfacial stacking and the dislocation network (yellow) in a representative small twist TGB ($\theta=1.61^{\circ}$) are shown in Fig.~\ref{figS:dislocation}.
The dislocation network for the low-load ($P=0$) undistorted case is triangular and remains 2D; while that for the high load ($P=3.4$ GPa) distorted TGB shows an extended 3D structure: 
The dislocations around the mini domains (MDs) penetrate into the layer beneath, causing frictional locking.

\section{Disorder parameter under $T=0$ loading/unloading}

\begin{figure*}[ht!]
\centering
\includegraphics[width=0.8\linewidth]{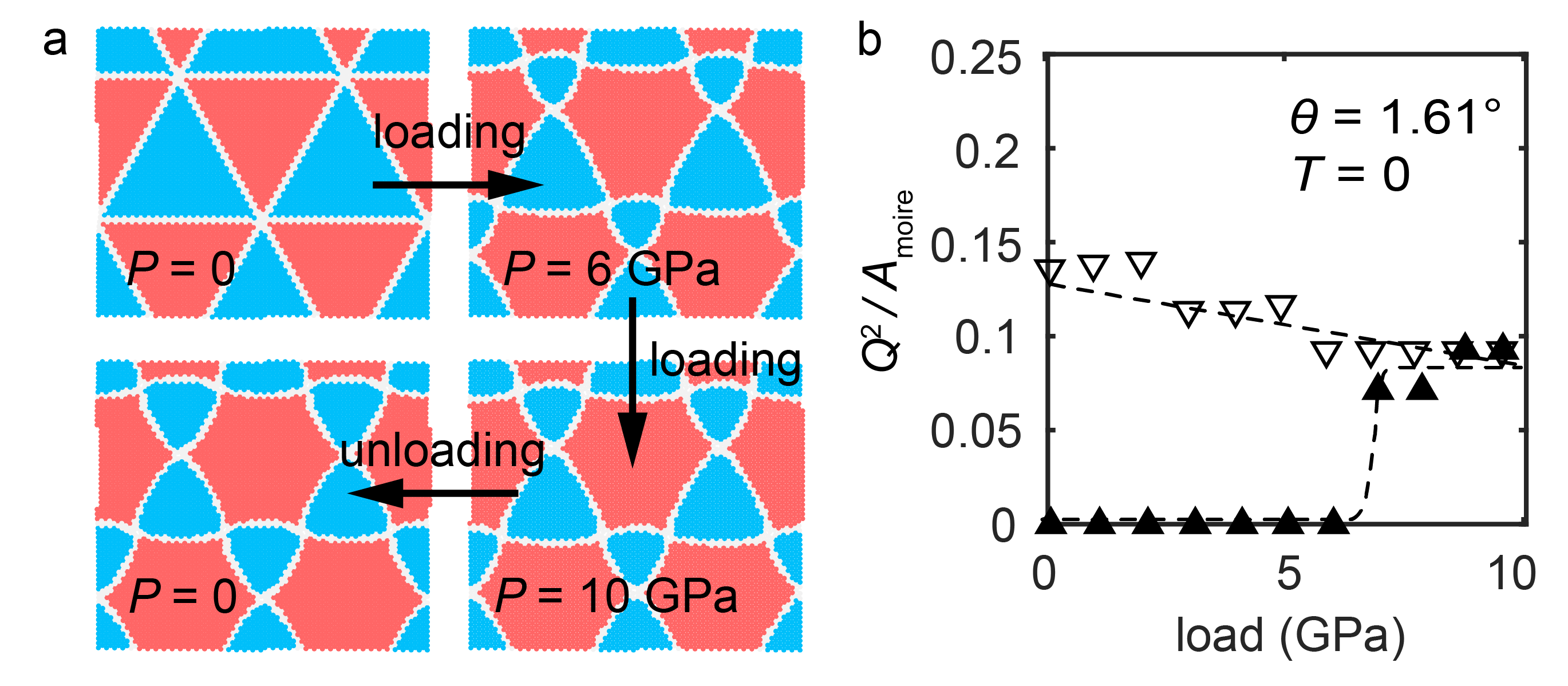}
\caption{a. Evolution of the interfacial stacking during load and unloading.
b. Dependence of disorder parameter $Q$ on load. 
Solid upward triangles: loading; hollow downward triangles: unloading.}
\label{figS:disorder}
\end{figure*}

In Fig.~4 of the main text and its associated discussions, we presented the evolution of the disorder parameter $Q$ during loading and unloading at room temperature $T=300$ K. 
As a supplement to the quasi-static loading and unloading results at 0 K (Fig.~2 of main text), we show in Fig.~\ref{figS:disorder}a the structural evolution and in Fig.~\ref{figS:disorder}b the dependence of $Q$ on the load, for a low-twist $\theta=1.61^{\circ}$ Au(111) TBG.
During loading, a jump of $Q$ appeared at $P\approx 6$ GPa.
Upon unloading, the distorted structure did not spontaneously revert back to undistorted.

\section{Up-down configurations}

\begin{figure*}[ht!]
\centering
\includegraphics[width=1.0\linewidth]{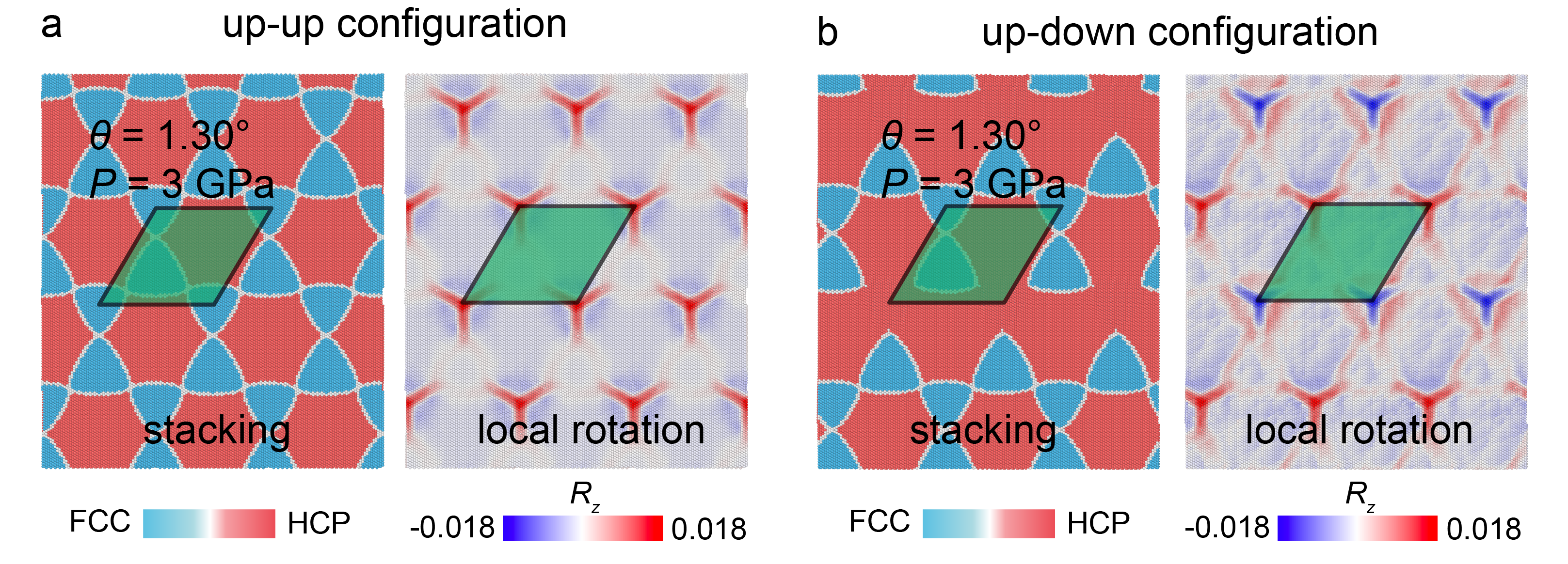}
\caption{Stacking and local rotation of the quasi-statically relaxed (a) ``up-up'' and (b) ``up-down'' configurations. Only one TGB layer is shown, the other is complementary. The green rhombus represents a single moir\'e unit cell.}
\label{figS:updown}
\end{figure*}

As the moir\'e size expands at decreasing $\theta$, the AA nodes move apart and their mutual interaction weakens.
At very small twists such as $\theta=1.30^{\circ}$, a novel ``up-down'' distorted configuration, appeared  upon loading, with about the same energy as the regular ``up-up'' configuration found at larger twists.
Up-up refers to the ``regular" distorted structure where all MDs form on the same layer of the TGB interface (Fig.~\ref{figS:updown}a). 
In the up-down structure instead, the MDs alternatively form on both layers of the TGB interface (Fig.~\ref{figS:updown}b). At the smallest twist angles both up-down and up-up configurations appeared with certain probabilities in several independent room temperature loading simulations.  

Although the two configurations do not significantly affect the transition pressure during loading -- both occur at $\approx 3.5$ GPa -- they have distinct differences.
The up-down configuration has a different symmetry, the in-plane moir\'e periodicity shifting from $(1\times 1)$ to $(2\times 1)$.
In addition, in the up-up configuration the local rotations at the mini-domains (originally AA nodes) are all aligned in the same direction, whereas in the up-down case, they are oppositely oriented (Fig.~\ref{figS:updown}).

After thermal equilibration at constant pressure $P=3$ GPa and annealing to 0 K for both configurations, we found that the potential energy of the up-up configuration is still slightly lower than that of the up-down configuration:
$\Delta E = E_\mathrm{up/down} - E_\mathrm{up/up} \approx 37.5$ meV for nearly $4\times 10^5$ atoms, an extremely small difference.
When we subsequently heated both systems from 0 K to room temperature (within 2 ns), neither spontaneously transferred into the other, suggesting a significant barrier between the two stable states.

\section{Critical twist angle and critical load}

Start with the extended Landau-type free energy of main text,
\begin{equation} 
 F(Q) = F_1 f(Q/a_0) + F_2 [1-f(Q/a_0)]
\end{equation}
the two terms
\begin{equation}
\begin{aligned}
    F_1 &= 3 E_\mathrm{AA} = 3E_0 +3\alpha P a^3 \\
    F_2 &= 3 E_\mathrm{vac} a Q^{-1} + \frac{1}{2} \tau_\mathrm{s} a^2 Q + \frac{1}{2} G \theta^2 a Q^{2}
\end{aligned}    
\end{equation}
describe the two competing states -- undistorted and distorted.

Here, $E_\mathrm{AA}$ is the excess energy per atom of the three central 
atoms of the undistorted AA node, $\alpha$ is a dimensionless parameter, $P$ is the normal load (units of pressure), $E_\mathrm{vac}$ is a corner vacancy formation energy, $\tau_\mathrm{s}$ is the shear strength, $G$ is the shear modulus, $\theta$ is the twist angle, $f$ is a cutoff function that is taken as $f(x) = e^{-x^2}$, and $a_0 = 3a/2$.

Focusing on the distorted state energy $F_2$, the energy minimum is obtained when $\partial F_2/\partial Q=0$, which gives one positive solution
\begin{equation}
    Q_0 = \frac{1}{6G\theta^2} \left( c_0^{1/3}+\frac{\tau^2 a^2}{c_0^{1/3}}-\tau a \right)
\end{equation}
where
\begin{equation}
\begin{aligned}
    c_0 &= (\sqrt{162c_1-c_2}+\sqrt{162c_1})^2 \\
    c_1 &= E_\mathrm{vac} G^2 \theta^4 \\
    c_2 &= \tau^3 a^3
\end{aligned}
\end{equation}
At a crossover twist angle
\begin{equation}
    \theta_0 = \left( \frac{\tau^3 a^3}{162 E_\mathrm{vac} G^2} \right)^{1/4}
\end{equation}
this solution stops. Therefore\\

(i) for $\theta > \theta_0$,
\begin{equation}
    \frac{162c_1}{c_2} = (\frac{\theta}{\theta_0})^4 \gg 1
\end{equation}
Thus, $c_0 \approx 648 c_1$, and Eq.~S3 
reduces to
\begin{equation}
    Q_0 \sim \left( \frac{3 E_\mathrm{vac}}{G \theta^2} \right)^{1/3}
\end{equation}
Substituting $Q_0$ back into $F_2$, 
the critical twist angle can be determined by solving $\partial ^2 F_2/\partial \theta^2 =0$:
\begin{equation}
    \theta_c = \left( \frac{125\tau^3 a^3}{81 E_\mathrm{vac} G^2} \right)^{1/4} \approx 4\theta_0
\end{equation}

With these $\theta_c$ and $Q_0$, the critical transition load can be further estimated by $F_1 = F_2$, which gives
\begin{equation}
    P_{tc} \approx \frac{1}{\alpha a^3} \left[ 3\sqrt{E_\mathrm{vac} \tau a^3/5} -E_0 \right]
\end{equation}\\

(ii) for $\theta<\theta_0$, $F_2$ can be simplified  to 
\begin{equation}
    F_2 = 3 E_\mathrm{vac} a Q^{-1} + \frac{1}{2} \tau_\mathrm{s} a^2 Q
\end{equation}
The energy minimum is  at 
\begin{equation}
    Q_0 \sim \left( \frac{6E_\mathrm{vac}}{\tau a} \right)^{1/2}
\end{equation}
with the transition load
\begin{equation}
    P_t \sim \frac{1}{\alpha a^3} \left[ \sqrt{2E_\mathrm{vac} \tau a^3/3} -E_0  \right] < P_{tc}
\end{equation}

\begin{figure*}[hb!]
\centering
\includegraphics[width=0.9\linewidth]{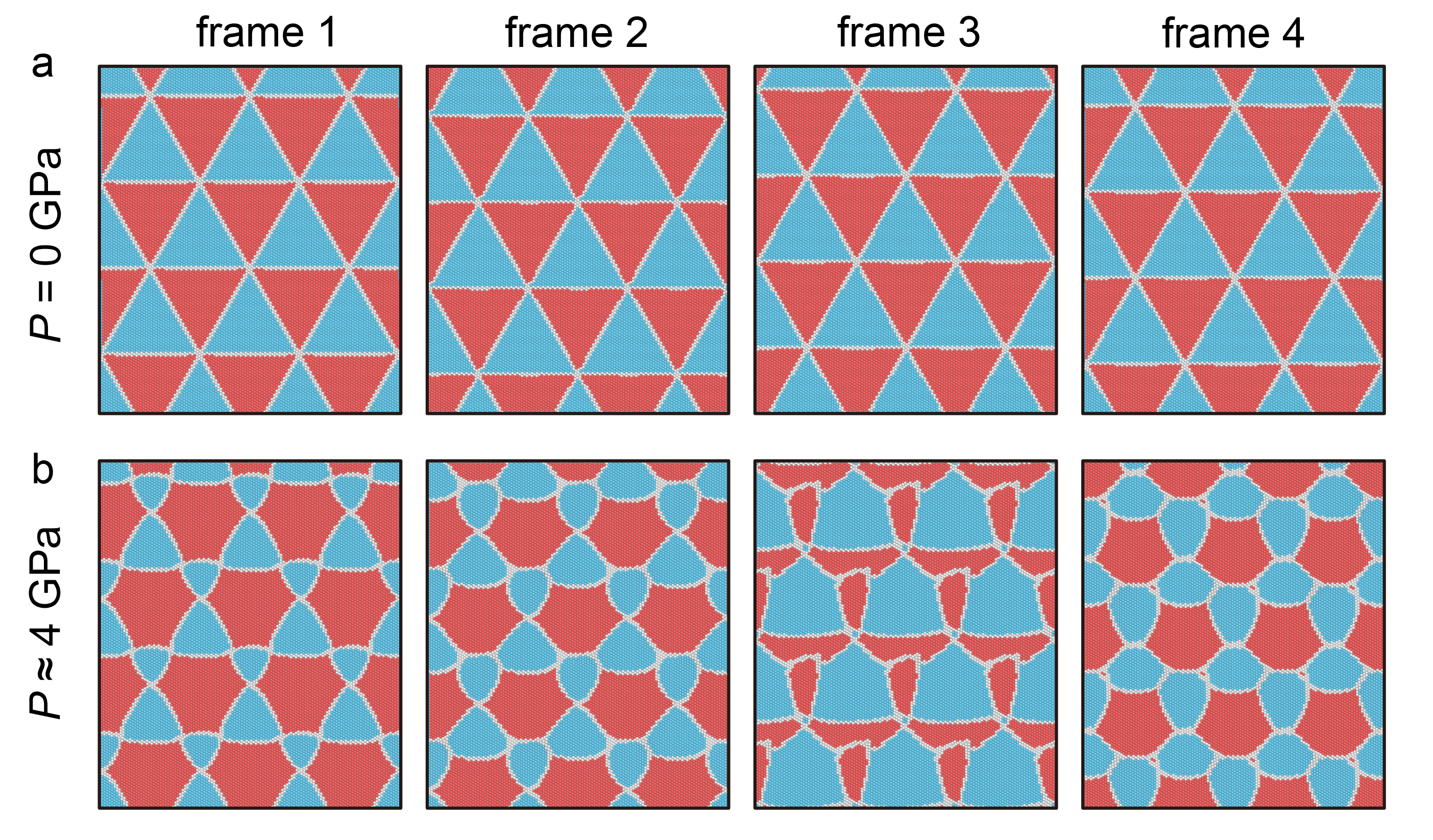}
\caption{Interfacial stacking of four representative frames during the quasi-static sliding simulation ($T=0$, $v_0 \to 0$) of a small twist TGB ($\theta=1.61^{\circ}$). 
a. Zero load $P=0$ case. b. High load $P=4$ GPa case.}
\label{figS:static}
\end{figure*}

\begin{figure*}[ht!]
\centering
\includegraphics[width=0.9\linewidth]{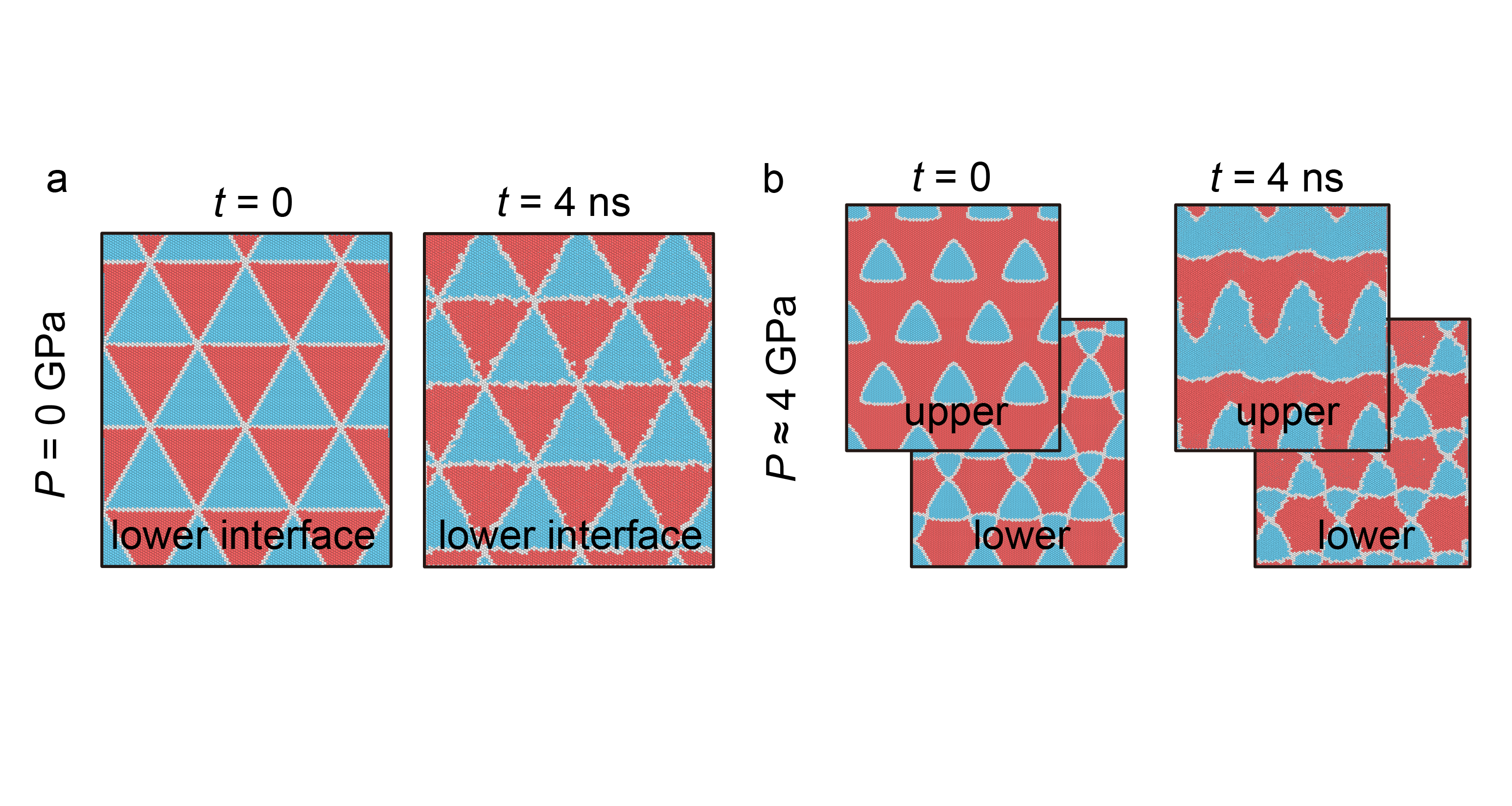}
\caption{Interfacial stacking before and after the dynamic sliding simulation ($T=300$ K, $v_0 = 10$ m/s) of a $\theta=1.61^{\circ}$ TGB.
a. Zero load $P=0$ case. b. High load $P=4$ GPa case.}
\label{figS:dynamic}
\end{figure*}

\section{Structural evolution during sliding}

\subsection{Zero load and high load}

In this section we present the evolution of the interfacial structure during the sliding process in both quasi-static (Fig.~\ref{figS:static}) and room temperature dynamic (Fig.~\ref{figS:dynamic}) simulations.

The quasi-static sliding simulations mimics the zero velocity, adiabatic limit of sliding at $T=0$. Four representative frames during the simulation are shown in Fig.~\ref{figS:static}.
At zero load, the equilibrium moir\'e pattern remains intact in the course of continuous, ideal, smooth ``superlubric" sliding. 
At high load by contrast, the strongly distorted moir\'es pattern undergoes periodic mechanical instabilities, suggestive of potential stick-slip sliding.

When real kinetic sliding simulations are carried out at room temperature (typically with $v_0 \sim $ m/s), both the motion and the friction are indeed superlubric in low load TGBs, consistent with the quasi-static picture -- the moir\'es smoothly slide remaining intact as in Fig.~\ref{figS:dynamic}a.
In the high load TGB, however, the real frictional evolution differs from the quasi-static expectations. As shown in Fig.~\ref{figS:dynamic}b where the $P=4$ GPa distorted TGB is shown, real sliding leads to irreversible structural changes especially in the  the upper layer. 
The initially locked TGB is irreversibly modified and damaged by plastic deformations. The nature of sliding is moreover different depending on velocity. 
At high velocity, the sliding occurs at the TGB interface, where the plastic deformation remains confined. At sufficiently low velocity and intermediate twist angles around $\theta_0$ however, another surprising regime appears, where the TGB interface does not slide, and shear takes place elsewhere in the crystal.
The MD provide in this case a locking of the TGB interface stronger than that among consecutive perfect (111) planes! \\

\begin{figure*}[ht!]
\centering
\includegraphics[width=0.9\linewidth]{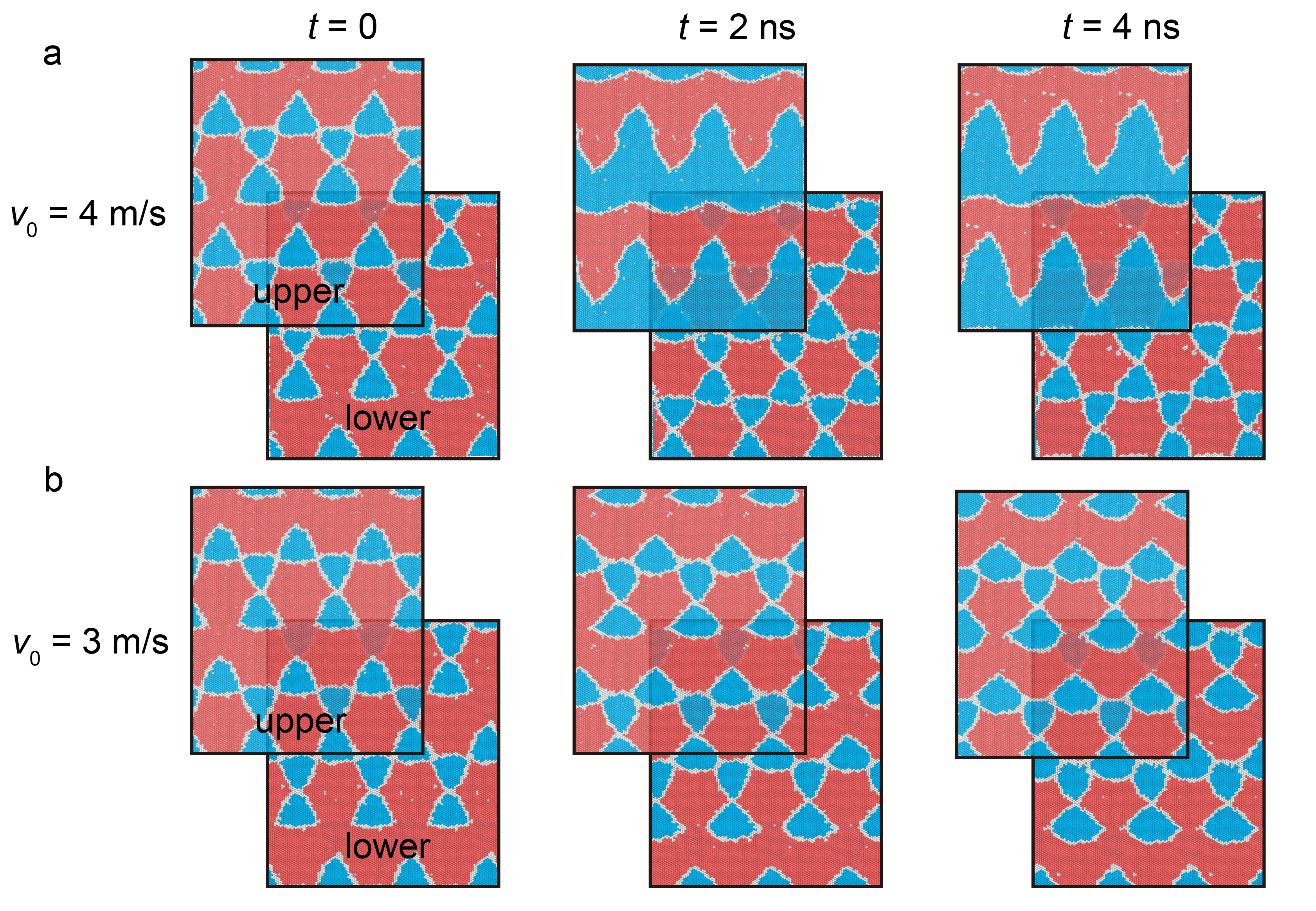}
\caption{Interfacial stacking of a $\theta = 1.61^{\circ}$ TGB under high load $P=4$ GPa ($T=300$ K, ``up-down'' at the beginning). 
(a) Plastic sliding regime in a representative high velocity case, $v_0=4$ m/s. (b) Interfacial locking regime in a low velocity case, $v_0=3$ m/s. 
Note that the transition velocity between two regimes is between 3 m/s and 4 m/s in our simulations.}
\label{figS:locking}
\end{figure*}

\subsection{Zero load: before loading and after unloading}

The unloading of the high pressure distorted TGB preserves the distortion in a metastable state, and that even at room temperature (Fig.~S7a). 
In that case however, a brief successive sliding readily promotes the TGB return to the undistorted state, thus restoring all superlubric features (Fig.~S7).

\begin{figure}[ht!]
\centering
\includegraphics[width=0.9\linewidth]{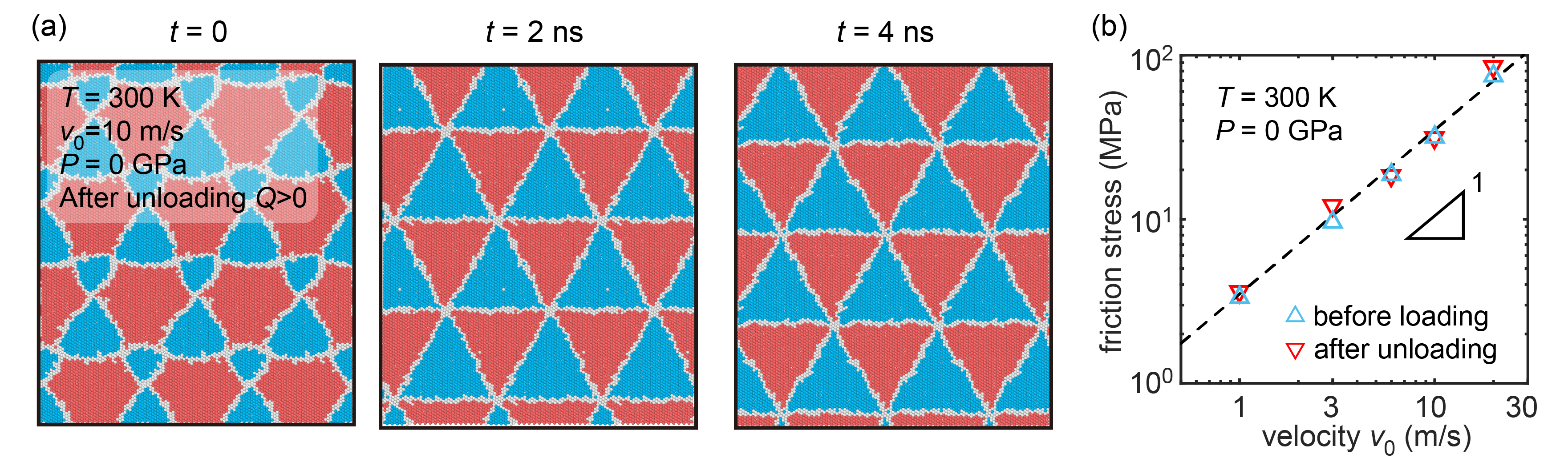}
\caption{Shear sliding of a small angle Au(111) TGB ($\theta=1.61^{\circ}$).
(a) Evolution of the structure at the lower interface.
(b) Velocity dependence of friction at zero load. Blue triangles: undistorted TGB, never loaded; red inverted triangles: after loading and successive unloading. The dashed line is a linear fit.} 
\label{figS:linear_v0}
\end{figure}

\section{Force field tests and Lennard-Jones fcc TGBs}

Two important questions are: (i) would a different parametrization of gold's interatomic forces yield similar or different results? 
(ii) would TGBs in other materials behave similarly or differently?
To address question (a) we performed additional quasi-static loading simulations on Au(111) TGBs with a very different force field. To address (b) we repeated simulations for Lennard-Jones (LJ)  fcc crystal TGBs ($\theta=1.61^{\circ}$). Both followed the same  protocol described in the main text.
For Au(111) TGBs  we used instead of Foiles potential, the alternative and very different Gupta potential \cite{Gupta.prb.1981,Rosato.phimag.1989},  also adopted in nanoscale gold cluster simulations \cite{Baletto.jcp.2002,Settem.smallsci.2024}.
For the LJ TGBs, we chose the same lattice constant as gold, i.e., $a=4.08~\mathrm{\AA}$, while the 12-6 LJ parameters were arbitrarily set as 
$\sigma=2.645~\mathrm{\AA}$ and $\varepsilon = 5$ meV, with a cutoff $15~\mathrm{\AA}$.

\begin{figure*}[ht!]
\centering
\includegraphics[width=0.8\linewidth]{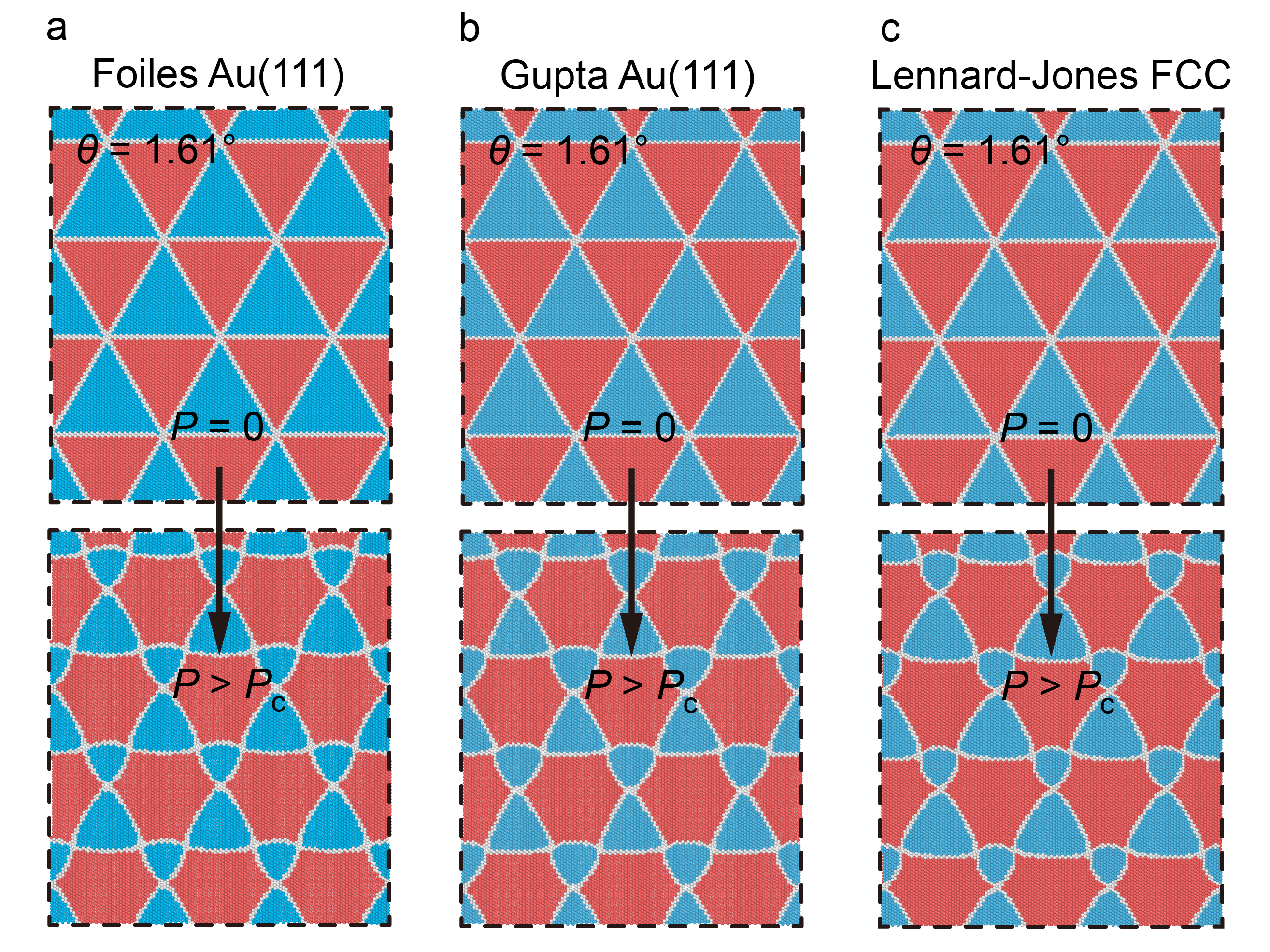}
\caption{Comparison of the interfacial stacking at zero ($P=0$) and high normal load ($P>P_c$) for Au(111) TGBs with Foiles force field (a) and Gupta force field (b), and for the LJ fcc TGB (c).}
\label{figS:lj}
\end{figure*}

Upon quasi-static loading, we consistently found a similarly relaxed undistorted moir\'e pattern at low loads (Fig.~\ref{figS:lj} upper panel).
As the load gradually grew we found the same distortion, again at $P_\mathrm{load} \approx 6$ GPa for Gupta's parametrization of the  Au(111) TGB;  and $P_\mathrm{load}\approx 0.5$ GPa for the LJ TGB). 
In both, the AA nodes transforms into commensurate mini-domains (Fig.~\ref{figS:lj} lower panel) -- consistent with the main text results. 
The structural transition and its mechanism appear therefore ubiquitous, and likely to be a general  property of 3D materials TGBs.

\bibliography{ref}